\documentclass[12pt]{iopart}

\usepackage{graphicx}
\usepackage{color}

\begin{document}

\title[Statistical distribution of the electric field driven switching]{Statistical distribution of the electric field driven switching of the Verwey state in Fe$_{3}$O$_{4}$}

\author{A A Fursina$^{1}$\footnote{Present address:  Department of Chemistry, University of Nebraska-Lincoln, 525 HAH, Lincoln, NE 68588-0304, USA}, R G S Sofin$^{2}$, I V Shvets$^{2}$, and D Natelson$^{3,4}$ }
\address{$^{1}$Department of Chemistry, Rice University, 6100 Main St., Houston, TX 77005, USA}
\address{$^{2}$CRANN, School of Physics, Trinity College, Dublin 2, Ireland}
\address{$^{3}$Department of Physics and Astronomy, Rice University, 6100 Main St., Houston, TX 77005}
\address{$^{4}$Department of Electrical and Computer Engineering, Rice University, 6100 Main St,.Houston, TX 77005}

\date{\today}


\begin{abstract} 
The insulating state of magnetite (Fe$_{3}$O$_{4}$) can be disrupted by a sufficiently large dc electric field.  Pulsed measurements are used to examine the kinetics of this transition.  Histograms of the switching voltage show a transition width that broadens as temperature is decreased, consistent with trends seen in other systems involving ``unpinning'' in the presence of disorder.  The switching distributions are also modified by an external magnetic field on a scale comparable to that required to reorient the magnetization.
\end{abstract}
\submitto{\NJP}
\maketitle

\section{Introduction}

Magnetite is an archetypal strongly correlated transition metal oxide, with properties not well described by single-particle band structure.  Below 858~K, magnetite, which may be written as Fe$^{3+}_{\mathrm{A}}$(Fe$^{2+}$Fe$^{3+}$)$_{\mathrm{B}}$O$_{4}$, is ferrimagnetically ordered, with the A and B sublattices having oppositely directed magnetizations.  The moments of the five unpaired $3d$ electrons of the tetrahedrally coordinated A-site Fe$^{3+}$ ions are compensated by those of the octahedrally coordinated B-site Fe$^{3+}$ ions.  The net magnetization results from the octahedrally coordinated B-site Fe$^{2+}$ that have four unpaired $3d$ electrons \cite{McQueeney:2005}.  Upon cooling, bulk magnetite undergoes a first-order phase transition from a moderately conducting high temperature state to a more insulating low temperature state at what is now called the Verwey\cite{Verwey:1939} temperature, $T_{\mathrm{V}}\approx 122$~K.  The change in electronic properties is coincident with a structural transition from a high temperature cubic inverse spinel to a low temperature monoclinic unit cell.  The nature of the ordered insulating state remains an active topic of current research\cite{Rozenberg:2006,Piekarz:2006,Schlappa:2008,Subias:2009}.  Experiments indicate the onset of multiferroicity\cite{Rado:1975} in magnetite below 40~K\cite{Alexe:2009}, further highlighting the rich physics in this correlated system.

Recently, nanostructured electrodes have been used to apply strong electric fields in the plane of magnetite films\cite{Lee:2008,Fursina:2008}.  Below $T_{\mathrm{V}}$, a sufficient applied voltage triggers a breakdown of the comparatively insulating low-temperature state and a sudden increase in conduction \cite{Lee:2008,Fursina:2008}.  This is an example of electric field-driven breakdown of a gapped state in strongly correlated oxides\cite{Asamitsu:1997,Oka:2005,Sugimoto:2008} similar to Landau-Zener breakdown in classic semiconductors.  The electric field-driven transition in magnetite is consistent with expectations\cite{Sugimoto:2008} based on such a mechanism  (via geometric scaling \cite{Lee:2008,Fursina:2008}, lack of intrinsic hysteresis \cite{Fursina:2009}, changes of both contact and bulk resistance at the transition \cite{Fursina:2010a,Fursina:2010b}).  These prior experiments examined films of various thicknesses, from 30~nm to 100~nm.  No strong thickness dependence was observed in the switching properties, consistent with the applied lateral electric field at the sample surface acting as the driver of the breakdown (though thinner films showed a less pronounced Verwey transition in low-bias resistance vs. temperature measurements, consistent with expectations).

Here we report studies of the statistical variations of this electric field-driven transition in Fe$_{3}$O$_{4}$, as a function of temperature and magnetic field perpendicular to the film surface (out-of-plane).  We find that there is a statistical distribution of switching voltages, $V_{\mathrm{SW}}$, that becomes more broad and shifts to higher voltages as $T$ is reduced.  We discuss these trends in the context of switching kinetics in other systems that exhibit similar trends.  The application of a magnetic field perpendicular to the plane of the Fe$_{3}$O$_{4}$ film alters the $V_{\mathrm{SW}}$, shifting the mean by several mV (several percent) and changing its shape, within a range of fields comparable to that required to reorient the magnetization out of plane.

\section{Experimental Techniques}

\begin{figure}[b]
\begin{center}
\includegraphics[clip,width=10cm]{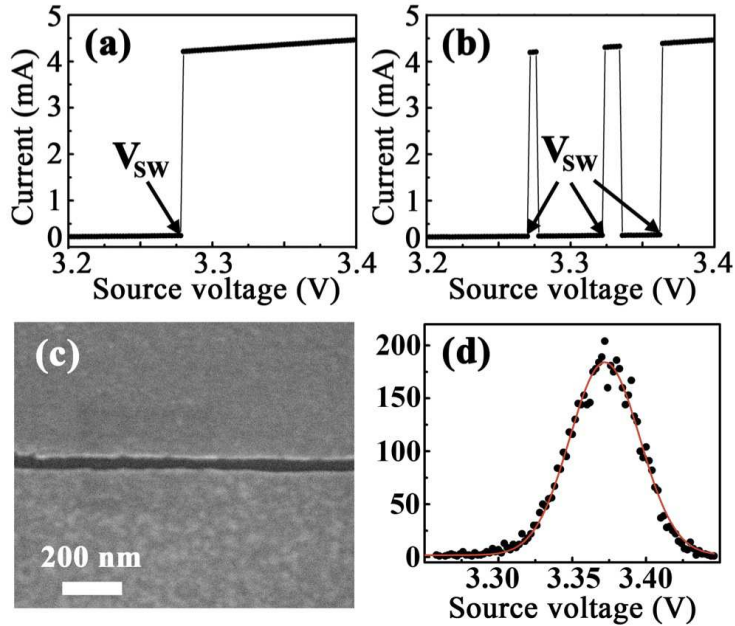}
\end{center}
\vspace{-5mm}
\caption{\small (color online) Details of $V_{\mathrm{SW}}$ distribution experiment. (a and b) The fragments of $I$-$V$ curves in the vicinity of a transition demonstrating one (a) and three (b) switching events in a single pulsed $I$-$V$ cycle. (c) Typical SEM image of Ti/Au electrodes, patterned on magnetite film surface, separated by nanogap $<$ 100 nm. (d) An example of $V_{\mathrm{SW}}$ distribution histogram at 90~K.}
\label{fig1}
\end{figure}

The 50~nm Fe$_{3}$O$_{4}$ (100) thin films used in the present study were grown on (100) oriented MgO single crystal substrates as described elsewhere  \cite{Shvets_backscat,Shvets_high_res_Xray}.   Contact electrodes (2nm adhesion layer of Ti and 15 nm layer of Au) were patterned by e-beam lithography on the surface of the Fe$_{3}$O$_{4}$ film.  As before \cite{Lee:2008,Fursina:2009}, $V_{\mathrm{SW}}$ scales linearly with the channel length, $L$ (the electrode spacing), implying an electric field-driven transition.  Long channels ($L >$ 100~nm) required large switching voltages that would alter the electrode geometry over numerous switching cycles, distorting the shape of $V_{\mathrm{SW}}$ histograms.  To minimize $V_{\mathrm{SW}}$, electrodes separated by 10-30~nm were patterned using a self-aligned technique \cite{Fursina:2008}.  Electrical characterization of the devices was performed  using a semiconductor parameter analyzer (HP 4155A).  To minimize self-heating when in the conducting state, the voltage was applied as pulses 500~$\mu$s in duration with a 5~ms period \cite{Fursina:2009,Fursina:2010a}.   The samples were cooled below $T_{\mathrm{V}}$ with no magnetic field applied, and the distribution of $V_{\mathrm{SW}}$ was obtained by executing several thousand consecutive forward pulsed $I$-$V$ sweeps in the vicinity of the transition point (typically a 0.2-0.3~V range) at a fixed (to within 50~mK) temperature, and recording the number of switching events at each voltage.  

Each voltage value is essentially an independent test to see if switching takes place under the pulse conditions.  Hence, some sweeps show one (figure~\ref{fig1} a) or several (figure~\ref{fig1} b) switching events.   Even if the system is switched to the conducting state at $V_{\mathrm{SW}}$(1), it may return to the Off state between pulses, and then switch to the On state at some higher voltage, $V_{\mathrm{SW}}$(2), and so on.  The $V_{\mathrm{SW}}$ distribution at a particular temperature is built by recording all switching events over several thousands (3000-6000) of $I$-$V$ cycles and then counting the number of switchings at a certain $V_{\mathrm{SW}}$, to produce a ``\# of counts'' vs. $V_{\mathrm{SW}}$ histogram.  A typical $V_{\mathrm{SW}}$ distribution at 90K is shown in figure~\ref{fig1} d.  The distribution is a single peak, symmetrical around the most probable $V_{\mathrm{SW}}$ value.  

\section{Results and Discussion}

\begin{figure}[b]
\begin{center}
\includegraphics[clip,width=10cm]{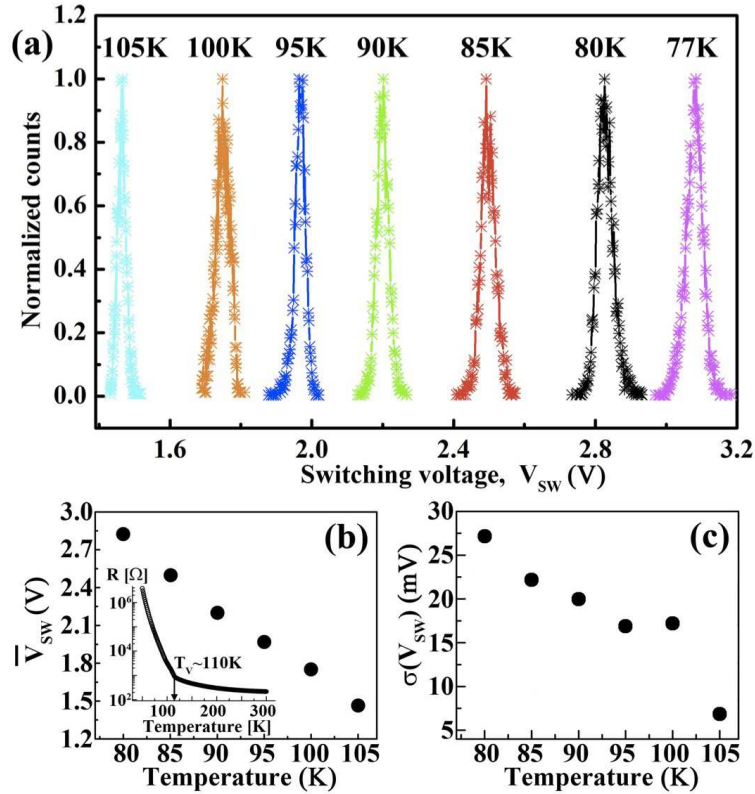}
\end{center}
\vspace{-5mm}
\caption{\small (color online) (a) Normalized $V_{\mathrm{SW}}$ distributions at different temperatures (77~K - 105~K). (b) Temperature dependence of the mean switching voltage, $\bar{V}_{SW}$. Inset shows zero-bias $R$ vs $T$ plot demonstrating $T_{\mathrm{V}} \sim$~110~K.(c) Temperature dependence of $V_{\mathrm{SW}}$ distribution width, $\sigma(V_{\mathrm{SW}})$ (black circles).  }
\label{fig2}
\vspace{-5mm}
\end{figure}

This procedure was repeated at each temperature below $T_{\mathrm{V}}$ ($\sim$~110K for devices under test; see figure~\ref{fig2}b inset), down to $\sim$ 75~K.  At $T < 75$~K, the high values of $V_{\mathrm{SW}}$ necessary led to irreversible alteration of the electrode geometry, resulting in asymmetric, distorted $V_{\mathrm{SW}}$ histograms.   Near 80-90~K, $I$-$V$ cycles with multiple $V_{\mathrm{SW}}$ events (figure~\ref{fig1}b) were observed more frequently. Thus, the total number of switching events observed varied with $T$, even with a fixed number of $I$-$V$ cycles at each temperature.  To compare $V_{\mathrm{SW}}$ distributions at different temperatures, the distributions were normalized, plotted as (\# of counts)/(max \# of counts) vs $V_{\mathrm{SW}}$, where ``max \# of counts'' is the number of events at the most probable $V_{\mathrm{SW}}$ and ``\# of counts'' is the number of events at a certain $V_{\mathrm{SW}}$.  Figure~\ref{fig2}a is an example of normalized $V_{\mathrm{SW}}$ distributions in the 77~K-105~K temperature range.  The measured widths of the $V_{\mathrm{SW}}$ distributions are not limited by temperature stability.

As has been discussed elsewhere\cite{Fursina:2009}, the use of pulses is essential to minimize the role of self-heating once the system has been driven into the more conducting state.  This self-heating and the short timescale\cite{Fursina:2009} required to raise the local temperature in the channel significantly makes it extremely challenging to determine directly whether the initial breakdown takesplace through the formation of a conducting filament or through a uniform switching; once a highly conducting path is formed, the whole channel rapidly becomes conducting through self-heating.   The filamentary picture is certainly likely, based on other breakdown phenomena in solids, and the statistical variation in $V_{\mathrm{SW}}$ is consistent with the idea of a process involving run-to-run variability associated with \textit{local} details rather than global material properties, but this is not definitive.

The $V_{\mathrm{SW}}$ distribution at each temperature is characterized by two main parameters:  the mean switching value $\bar{V}_{\mathrm{SW}}=(\sum_{i=1}^N V_{{\mathrm{SW}},i})/N$, where $N$ is the total number of switching events; and the width of distribution, calculated as a standard deviation:
$\sigma(V_{\mathrm{SW}})=\sqrt{(\sum_{i=1}^N (V_{{\mathrm{SW}},i}-\bar{V}_{sw})^2)/{N-1}}$.  As expected, the $\bar{V}_{\mathrm{SW}}(T)$ has the same $T$-dependence (see figure~\ref{fig2} b) as $V_{\mathrm{SW}}(T)$ in single $I$-$V$ experiments described in previous publications \cite{Lee:2008,Fursina:2009}.  More interesting is the $\sigma(V_{\mathrm{SW}})$ temperature dependence, showing broadening of the $V_{\mathrm{SW}}$ distribution as the temperature decreases (figure~\ref{fig2} c).  We note a deviation from monotonous temperature dependence of $\sigma(V_{\mathrm{SW}})$ at 100~K, observed in several devices tested.   This is a temperature well below $T_{\mathrm{V}} = 110~K$ (see figure~\ref{fig2}b inset), where several physical parameters (resistance, heat capacity and magnetoresistance) change abruptly.     

This increase in $\sigma(V_{\mathrm{SW}})$ as temperature decreases is rather counter-intuitive.  One might expect ``freezing'' of temperature fluctuations and decrease in thermal noise the temperature decreases and, thus, narrowing of $V_{\mathrm{SW}}$ distributions.  The field-driven breakdown of the insulating state is an example of the ``escape-over-barrier'' problem, addressed generally by A.~Garg \cite{Garg:1995}.  Below $T_{\mathrm{V}}$, the (temperature-dependent) effective free energy of the electronic system is at a global minimum value in the insulating state, while the external electric field modifies the free energy landscape, lowering the free energy of another local minimum corresponding to the more conducting state.  As the external field is increased beyond some critical value, the minimum corresponding to the more conducting state becomes the global minimum.  The nonequilibrium transition to the conducting state then corresponds to some process that crosses the free energy barrier between these minima.  At  a sufficiently large value of the external field, the free energy has only one minimum, corresponding to the conducting state.  

As Garg showed, one may consider thermal activation over the free energy barrier as well as the possibility of quantum escape.  This free energy picture predicts a broadening of the transition driving force ($V_{\mathrm{SW}}$ here) distribution as the absolute value of the driving force increases, consistent with our observations.  This free energy picture has proven useful in studying other nonequilibrium transitions, such as magnetization reversal in nanoparticles \cite{Wernsdorfer:1997} and nanowires \cite{Varga:2003,Varga:2004}.    Pinning due to local disorder is one way to find increasing distribution widths as $T \rightarrow 0$, as seen in investigations of field-driven magnetization reversal in nanowires \cite{Varga:2003,Varga:2004}.  Unfortunately, quantitative modeling in this framework requires several free parameters and is difficult without a detailed understanding of the underlying mechanism.   

Qualitatively similar phenomenology (distribution of switching thresholds that broadens as $T$ is decreased) is also observed in the current-driven superconducting-normal transition in ultrathin nanowires\cite{Sahu:2009,Pekker:2009}.  In this latter case as in ours, self-heating in the switched state is of critical importance, as is the temperature variation of the local thermal path.  Again, quantitative modeling using this self-heating approach would require the introduction of multiple parameters that are difficult to constrain experimentally, as well as detailed thermal modeling of the nanoscale local effective temperature distribution, and is beyond the scope of this paper.

\begin{figure}[h]
\begin{center}
\includegraphics[clip,width=8.5cm]{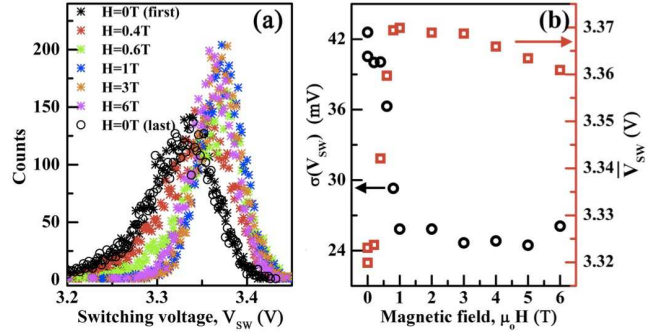}
\end{center}
\vspace{-5mm}
\caption{\small (a) Examples of $V_{\mathrm{SW}}$ distributions at selected magnetic fields (T=80~K).(b) Magnetic field dependence of the mean switching value, $\bar{V}_{\mathrm{SW}}$ (red squares), and the width of $V_{\mathrm{SW}}$ distributions, $\sigma(V_{\mathrm{SW}})$ (black circles).}
\label{Vsw_H_dep}
\end{figure}

We also examined the dependence of the switching distributions on applied out-of-plane magnetic field, $H$.  $V_{\mathrm{SW}}$ distributions (3000 cycles each) were collected consecutively at 12 magnetic field values:  
0~T (first) $\rightarrow$ 0.2~T $\rightarrow$ 0.4~T $\rightarrow$ 0.6~T $\rightarrow$ 0.8~T $\rightarrow$ 1~T $\rightarrow$ 2~T $\rightarrow$ 3~T $\rightarrow$ 4~T $\rightarrow$ 5~T $\rightarrow$ 6~T $\rightarrow$ 0~T(last).  Figure~\ref{Vsw_H_dep} a shows the resultant $V_{\mathrm{SW}}$ distributions at 80~K at several selected magnetic fields. As can be seen, magnetic field shifts the $V_{\mathrm{SW}}$ peak to higher $V$ values and narrows the $V_{\mathrm{SW}}$ distributions. To reassure that the observed $\bar{V}_{\mathrm{SW}}$ shift is not from irreversible changes in the device, a control experiment returning to $H=0$~T was performed after experiments in all non-zero magnetic fields.  The $V_{\mathrm{SW}}$ distributions at $H=0$~T initially [$H=0$~T (first)] and in the end [$H=0$~T (last)] are identical (see figure~\ref{Vsw_H_dep} a), meaning that observed changes in $V_{\mathrm{SW}}$ distributions (shift and narrowing) are indeed caused by the applied magnetic field.   Figure~\ref{Vsw_H_dep} b quantifies the dependence of $\bar{V}_{\mathrm{SW}}$ and $\sigma(V_{\mathrm{SW}})$.  It is clear that both parameters saturate as $H$ is increased beyond 1~T, {\it i.e.};  further increases of $H$ up to 6~T have no significant effect.   Magnetic field of the opposite polarity (not shown) has exactly the same effect on the position and the width of $V_{\mathrm{SW}}$ distributions.  Note that the shape of the distribution, in particular its asymmetry about the peak value of $V_{\mathrm{SW}}$, evolves nontrivially with magnetic field, becoming more symmetric in the high field limit.

\begin{figure}[h]
\begin{center}
\includegraphics[clip,width=8.0cm]{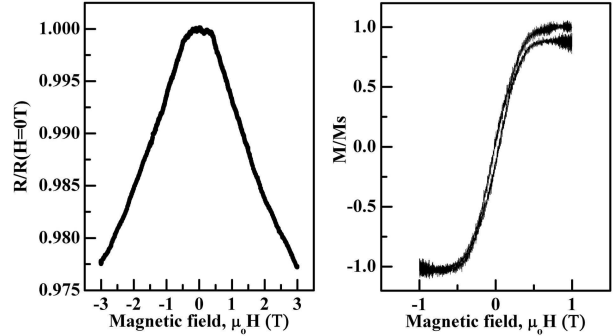}
\end{center}
\vspace{-5mm}
\caption{\small Dependences of the resistance ($R/R(H=0T)$) (a) and the magnetization ($M/M_{\mathrm{s}}$) (b) on the out-of-plane magnetic field applied.}
\label{R_M_vs_H}
\end{figure}

We consider whether this magnetic field dependence of the switching originates with some dependence of the bulk resistance or contact resistances, as this would alter the electric field distribution in the channel.   Since $V_{\mathrm{SW}}$ scales with the channel length, $L$  \cite{Lee:2008,Fursina:2009} {\it i.e.}, as does the resistance of the channel ($R \sim L$), one might expect an \textit{increase} in $V_{\mathrm{SW}}$ could result from an increase in the device resistance with applied magnetic field.  However, Fe$_3$O$_4$ has a negative magnetoresistance (MR)  \cite{Sofin:2005,deTeresa:2007,Eerenstein:2002}.   Figure~\ref{R_M_vs_H} a shows an example of normalized resistance dependence, $R/R(H=0T)$, on the out-of-plane magnetic field at 80~K.  The resistance remains effectively unchanged up to $\sim$0.4~T and then decreases as $|H|$ increases.  Thus, when $\bar{V}_{\mathrm{SW}}$ and $\sigma(V_{\mathrm{SW}})$ experience the predominance of their changes upon $H$ application ($H<1T$, see figure~\ref{Vsw_H_dep}), the resistance of the device either stays constant or decreases.  In the $H$ range when $R$ experiences significant changes (see figure~\ref{R_M_vs_H} a), $\bar{V}_{\mathrm{SW}}$ and $\sigma(V_{\mathrm{SW}})$ remain essentially unchanged (figure~\ref{Vsw_H_dep}). Therefore, the shift of $\bar{V}_{\mathrm{SW}}$ in the presence of $H$ does not originate from the change in the resistance value of the device.

Another Fe$_3$O$_4$ film parameter effected by $H$ is the magnetization of the film.  Figure~\ref{R_M_vs_H} shows the normalized out-of-plane magnetization, $M/M_{\mathrm{s}}$, as a function of the out of plane $H$, where $M$ is the magnetization of the film and $M_{\mathrm{s}}$ is the saturated magnetization.  This data is consistent with prior measurements on magnetite films\cite{Zhou:2004}  While we do not know the microscopic arrangement of $\mathbf{M}$ in the film in the absence of an external $\mathbf{H}$, magnetostatic energy considerations mean that $\mathbf{M}$ under that condition lies in the plane of the film.   The $H$ range over which $M$ is fully reoriented out of the plane (up to 1~T) matches the $H$ range of changes in the position of $\bar{V}_{\mathrm{SW}}$ and $\sigma(V_{\mathrm{SW}},T)$ (fig.~\ref{Vsw_H_dep} b).   This suggests (though does not prove) that the switching kinetics parameters $\bar{V}_{\mathrm{SW}}$ and $\sigma(V_{\mathrm{SW}},T)$, and therefore the stability of the gapped, low temperature, insulating state is tied the magnetization direction of magnetite films.

This observation is intriguing because it is not clear how the nonequilibrium breakdown of the low temperature state would be coupled to the magnetization.  Possible factors include magnetoelastic effects such as magnetostriction\cite{Tsuya:1977} ($\sim$ parts in $10^4$ per Tesla) affecting the tunneling matrix element between $B$-site iron atoms; and spin-orbit coupling playing a similar role\cite{Yamauchi:2010}.   There have been reports of significant magnetoelectric and multiferroic effects in magnetite \cite{Rado:1975,Alexe:2009}, and a recent calculation \cite{Yamauchi:2010} argues that these originate through the interplay of orbital ordering and on-site spin-orbit interactions of the B-site electrons.  In this picture, reorientation of the spin distorts the partially filled minority-spin orbitals occupied on the B-site (formally) Fe$^{2+}$ ions.  Such a distortion would be a natural explanation for the observed correlation between $\mathbf{M}$ and the kinetics of the electric field-driven breakdown of the ordered state, which directly involves the motions of those charge carriers.  It is unclear how this kind of spin-orbit physics would explain the evolution of the $V_{\mathrm{SW}}$ distribution, however.  It would also be worth considering whether there is any correlation between the characteristics of the switching distributions reported here, and the recently observed glassy relaxor ferroelectric relaxations in bulk magnetite crystals\cite{Schrettle:2011}.

Additional, detailed experiments as a function of directionality of $\mathbf{H}$, $\mathbf{M}$, and crystallographic orientation should be able to test these alternatives.  With the existing (100) films, studies of  $\bar{V}_{\mathrm{SW}}$ and $\sigma(V_{\mathrm{SW}},T)$ as a function of $H$ in the plane as well as perpendicular to the plane should be able to access the tensorial form of the $H$ dependence.  Comparison with appropriately directed $M$ vs. $H$ data as a function of temperature would be a clear test of whether the observed agreement between $H$-field scales (in $V_{\mathrm{SW}}$ and reorientation of $\mathbf{M}$ is coincidental.  Further measurements on films grown with different crystallographic orientations would serve as a cross-check.  It is important to note, however, that the acquisition of such data is very time intensive due to the need to acquire many thousands of switching events.  In turn, there is a companion requirement of extremely good device stability, to avoid irreversible changes in the metal configuration over the thousands of switching cycles.

\section{Conclusions}

We have studied the statistical distribution of the electric field needed for breakdown of the low temperature state of Fe$_{3}$O$_{4}$.  The distribution of critical switching voltages moves to higher voltages and broadens, as $T$ is reduced.  This broadening is consistent with phenomenology in other nonequilibrium experimental systems incorporating disorder and thermal runaway effects.  The breakdown distributions are altered by modest external magnetic fields normal to the film, suggesting a need for further experiments to understand the connection between magnetization and breakdown of the correlated state.  

The authors acknowledge valuable conversations with Paul Goldbart and David Pekker.  This work was supported by the US Department of Energy grant DE-FG02-06ER46337.  DN also acknowledges the David and Lucille Packard Foundation and the Research Corporation.  RGSS and IVS acknowledge the Science Foundation of Ireland grant 06/IN.1/I91.

%
\end{document}